\documentclass[conference]{IEEEtran}

\ifdefined\labelindent
\else
\newlength{\labelindent}
\fi

\usepackage[utf8]{inputenc}
\usepackage[T1]{fontenc}
\usepackage{graphicx}
\usepackage{amsmath, amssymb, amsfonts}
\usepackage{booktabs, multirow, array, tabularx}
\usepackage{color, colortbl, xcolor}
\usepackage{stfloats}
\usepackage{soul}
\usepackage[final]{hyperref}
\usepackage{tikz}
\usepackage{cite}
\usepackage{algorithm}
\usepackage{algpseudocode}
\usepackage{textcomp}
\usepackage{titlesec} 

\titlespacing*{\section}{0pt}{1ex}{1ex}
\titlespacing*{\subsection}{0pt}{1ex}{1ex}

\definecolor{bestcell}{HTML}{D9EDF7}
\definecolor{lime}{HTML}{A6CE39}

\DeclareRobustCommand{\orcidicon}{%
    \begin{tikzpicture}
        \draw[lime, fill=lime] (0,0)
            circle [radius=0.16]
            node[white] {\tiny ID};
        \draw[white, fill=white] (-0.0625,0.095)
            circle [radius=0.007];
    \end{tikzpicture}%
    \hspace{0mm}%
}

\newcommand{\orcidauthor}[2]{#1\,\orcidicon}

\begin{document}

\title{Secure mmWave Beamforming with Proactive-ISAC Defense Against Beam-Stealing Attacks}

\author{
    \IEEEauthorblockN{
        Seyed Bagher Hashemi Natanzi\,\orcidauthor{}{0000-0003-1524-8669}\IEEEauthorrefmark{1},
        Hossein Mohammadi\,\orcidauthor{}{0009-0007-7642-6891}\IEEEauthorrefmark{2},
        Bo Tang\,\orcidauthor{}{0000-0001-5708-766X}\IEEEauthorrefmark{1},
        Vuk Marojevic\,\orcidauthor{}{0000-0002-1217-7052}\IEEEauthorrefmark{2}
    }
    \IEEEauthorblockA{
        \IEEEauthorrefmark{1}Electrical and Computer Engineering Department, Worcester Polytechnic Institute, USA\\
        \IEEEauthorrefmark{2}Electrical and Computer Engineering Department, Mississippi State University, USA\\
        Email: \texttt{\{snatanzi, btang1\}@wpi.edu}, \texttt{\{hm1125, vm602\}@msstate.edu}
    }
}

\maketitle

\begin{tikzpicture}[remember picture, overlay]
    \node[anchor=north, yshift=-0.2cm] at (current page.north) {%
        \fbox{\parbox{\dimexpr\textwidth-2\fboxsep\relax}{\centering\small%
        \color{red}This work has been submitted to the IEEE Conference for possible publication. Copyright may be transferred without notice, after which this version may no longer be accessible.}}};
\end{tikzpicture}

\vspace{-0.7cm}
\begin{abstract}

Millimeter-wave (mmWave) communication systems face increasing susceptibility to advanced beam-stealing attacks, posing a significant physical layer security threat. This paper introduces a novel framework employing an advanced Deep Reinforcement Learning (DRL) agent for proactive and adaptive defense against these sophisticated attacks. A key innovation is leveraging Integrated Sensing and Communications (ISAC) capabilities for active, intelligent threat assessment. The DRL agent, built on a Proximal Policy Optimization (PPO) algorithm, dynamically controls ISAC probing actions to investigate suspicious activities. We introduce an intensive curriculum learning strategy that guarantees the agent experiences successful detection during training to overcome the complex exploration challenges inherent to such a security-critical task. Consequently, the agent learns a robust and adaptive policy that intelligently balances security and communication performance. Numerical results demonstrate that our framework achieves a mean attacker detection rate of 92.8\% while maintaining an 
average user SINR of over 13 dB.

\enlargethispage{-1.5\baselineskip}
\end{abstract}

\IEEEoverridecommandlockouts
\vspace{1mm}

\IEEEoverridecommandlockouts
\vspace{1mm}

\IEEEpeerreviewmaketitle

\section{Introduction}
\label{sec:Introduction}

mmWave communications provide high data rates for applications like AR/VR and connected vehicles~\cite{10422712}. This is achieved via highly directional beamforming, which mitigates severe path loss but introduces physical layer vulnerabilities~\cite{10735492}. Beam-stealing attacks, where adversaries hijack or eavesdrop beams, threaten link integrity and confidentiality~\cite{10.1145/3212480.3212499}. Concurrently, ISAC is emerging as a key 6G feature, efficiently using shared resources for dual functionality~\cite{10536135}. ISAC's integration however also introduces new security threats.

Securing mmWave links against sophisticated beam-stealing attackers that may employ intelligent, adaptive strategies and exploit protocol knowledge to evade conventional defenses is a significant challenge. For instance, simple Power Delay Profile (PDP) analysis, although useful, can be circumvented by attackers capable of subtle manipulations~\cite{10.1145/3212480.3212499}. This underscores the urgent need for robust, proactive, and adaptive defense mechanisms capable of countering dynamic and intelligent threats to ensure a reliable and trustworthy user experience. 

\begin{figure}[t]
    \centering
    \includegraphics[width=0.7\columnwidth]{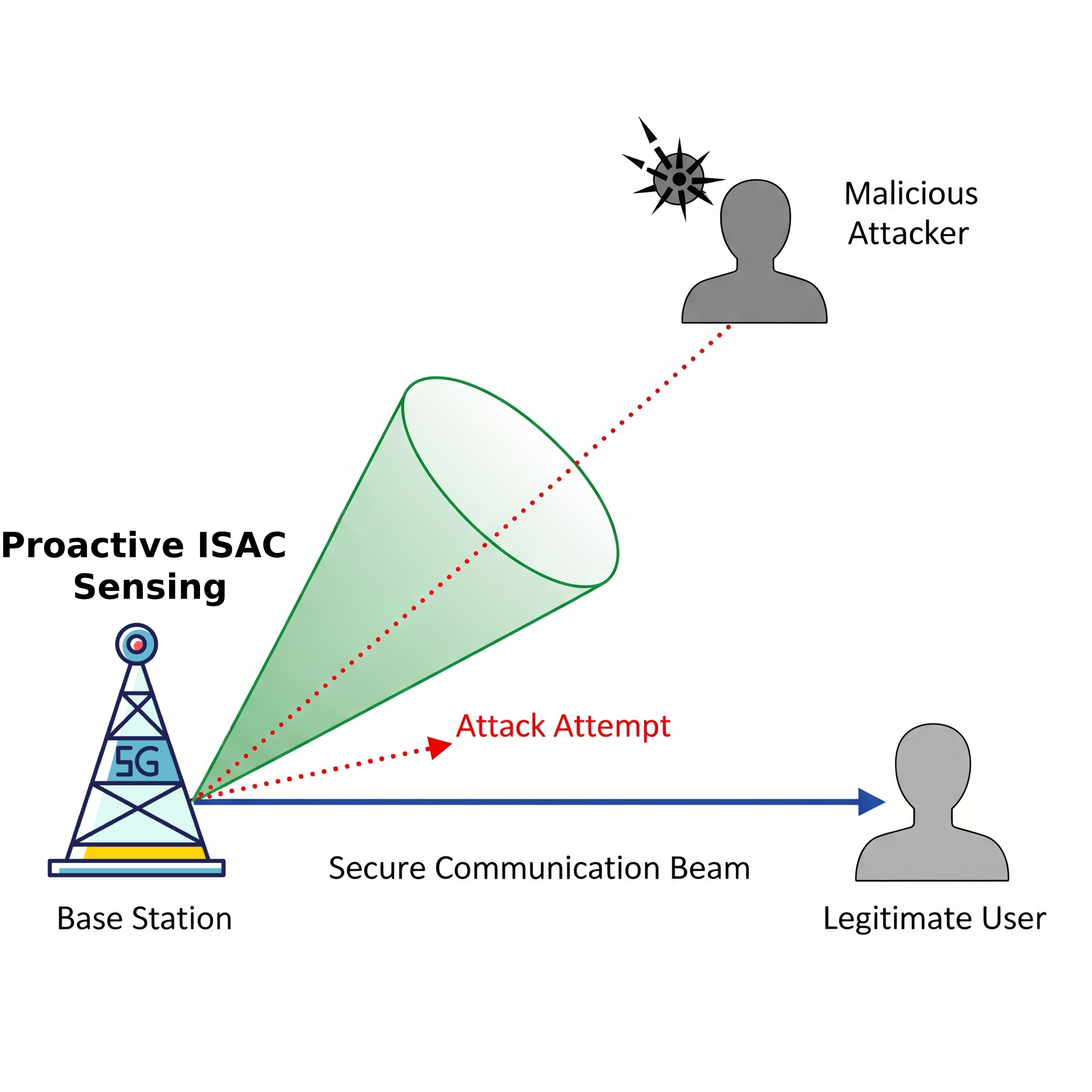}
     \vspace{-9mm} 
    \caption{{System and defense overview. The base station maintains a secure communication link with the legitimate user while facing a beam-stealing attempt from a malicious attacker. The integrated ISAC module actively probes the environment, detects potential threats, and informs the DRL agent, which dynamically adjusts the beam direction and sensing effort accordingly.}
}
    \label{fig:system_overview}
\end{figure}

Fig.~\ref{fig:system_overview} illustrates the threat and defense model considered in this work. A malicious attacker attempts to hijack the communication beam from a legitimate user. To counter this, we propose a novel framework where a DRL agent empowers the base station with proactive and adaptive beam control strategies. A key innovation is the use of ISAC not merely as a communication aid, but as an active sensing tool: the DRL agent leverages ISAC outputs to dynamically probe, detect, and respond to potential attacks. However, due to the complexity of the environment, standard agents often fail to discover secure policies. To address this, we introduce an intensive curriculum learning approach that ensures early successful detection experiences, guiding the agent toward robust convergence.

The primary contributions are: 
\begin{itemize}
    \item A security framework built on an advanced PPO agent for stable learning in a complex state-action space.
    \item A novel, intensive curriculum learning strategy that guarantees the agent experiences successful threat detection, a critical step to overcome the deep exploration challenges inherent in security-critical applications.
    \item A comprehensive analysis showing that the agent acquires a sophisticated, adaptive strategy that balances the detection rate with user communication quality. 
\end{itemize} 
This holistic approach significantly improves the resilience and security of mmWave links, ultimately enhancing the dependability of consumer oriented mmWave applications.

The remainder of this paper is organized as follows. Section~\ref{sec:related_work} reviews the related work. Section~\ref{sec:System_Model} introduces the system model, problem formulation, and overall framework. Section~\ref{sec:Proposed_Method_Content} presents the proposed DRL agent and its integration with ISAC. Section~\ref{sec:Results} provides the simulation setup and performance evaluation. Finally, Section~\ref{sec:Conclusion} concludes the paper.

\section{Related Work}\label{sec:related_work}

Beam-stealing attacks pose critical security threats in mmWave systems. Steinmetzer et al.~\cite{10.1145/3212480.3212499} practically demonstrate such attacks on IEEE 802.11ad networks: By injecting forged feedback into the sector sweep protocol, they enabled Man-in-the-Middle (MITM) relays for eavesdropping. Addressing beam-stealing attacks without reliance on training data, Yang et al.~\cite{Yang2022} propose an image processing methodology. Their approach uses a Received Signal Strength Indicator (RSSI) map for joint detection and localization of multiple attackers, achieving 91\% detection rates and sub-meter accuracy.
Li et al.~\cite{10735492} propose SecBeam to counter sophisticated beam-stealing, specifically amplify-and-relay (AnR) attacks that can circumvent beacon authentication. This protocol analyzes the PDP to detect manipulated signal paths by verifying that legitimate, shorter paths exhibit stronger and earlier-arriving signals compared to potentially amplified, longer relay paths. Recent active attacks such as BeamCraft~\cite{10.1145/3636534.3690669}, which successfully manipulate Wi-Fi traffic by injecting forged Beamforming Feedback Information (BFI), demonstrate the critical need for proactive defenses against such clear-text feedback vulnerabilities.

Further exploring physical layer security, Qiu et al.~\cite{10026317} address hybrid threats in mmWave environments by introducing an artificial noise (AN)-aided robust multi-beam secure communication scheme. Their work centers on the joint design of information and AN beamforming to counteract coexisting active jammers and passive eavesdroppers, considering imperfect adversary Channel State Information (CSI), where the legitimate receivers employ Minimum Variance Distortionless Response (MVDR) for jamming suppression. In the context of ISAC, Xu et al.~\cite{xu2025synergizingcoverttransmissionmmwave} combine physical layer covert transmission with ISAC functionalities. Their work designs transmit beamforming for both fully digital and hybrid architectures, enabling confidential, undetected communication to a covert user while concurrently supporting regular communication and sensing tasks, even under imperfect warden CSI. Furthermore, the security of Artificial Intelligence (AI) models for mmWave beamforming prediction is of critical concern. Kuzlu et al.~\cite{Kuzlu2023} highlight the susceptibility of deep learning based beam predictors to adversarial attacks that corrupt input data 
and explore adversarial training and defensive distillation as mitigation techniques.
While prior works focus on securing mmWave systems against attacks, optimizing performance in dynamic 6G environments is equally critical. Our recent work~\cite{natanzi2025onlinelearningbasedadaptivebeam} uses DRL for adaptive beam switching to improve throughput and SNR in dynamic 6G environments. Similarly, Mohammadi et al.~\cite{10773704} employ multipath communications to counter 5G jamming attacks, enhancing physical layer security against physical layer threats.

Our DRL framework uniquely employs ISAC for proactive defense against mmWave beam-stealing. It distinctively optimizes secure beamforming and ISAC probing for enhanced situational awareness and robust and adaptive countermeasures.
\section{System Model and Problem Formulation}
\label{sec:System_Model}
\subsection{System Model}
\label{subsec:System_Model_Content}

We consider a downlink mmWave ISAC scenario where a base station (BS) communicates with a legitimate user equipment (UE) while attempting to detect a potential beam-stealing attacker. The BS employs narrow directional beamforming to maximize link quality and simultaneously performs active sensing to identify suspicious targets in its environment.

Let $\mathbf{h} \in \mathbb{C}^N$ denote the channel vector between the BS and the UE, and let $\mathbf{w} \in \mathbb{C}^N$ represent the BS beamforming vector. The received signal is modeled as:
\[
y = \mathbf{h}^H \mathbf{w} x + n,
\]
where $x$ is the transmitted symbol and $n$ is complex Gaussian noise with variance $\sigma^2$.

The SINR observed by the UE is:
\[
\text{SINR} = \frac{|\mathbf{h}^H \mathbf{w}|^2}{\sigma^2}.
\]

The BS dynamically adjusts its beam direction and ISAC sensing effort. The sensing module models the detection probability of the attacker as a function of sensing effort $e_t$ and distance $d_t$:
\[
P_d = 1 - e^{-\alpha e_t} e^{-\beta d_t},
\]
where $\alpha, \beta > 0$ model sensing efficiency and distance attenuation.

Sensor measurements are subject to Gaussian errors in range and azimuth, and the agent observes a state vector $\mathbf{s}_t$ that includes communication and sensing features.

\subsection{Problem Formulation}
\label{subsec:Problem_Formulation_Content}

The primary objective is to devise a control policy, \(\pi(a_t|s_t)\), that prioritizes security by maximizing attacker detection while maintaining adequate communication services. This is formulated as a DRL problem where the agent learns to maximize the expected cumulative discounted reward:
\[
G_t = \sum_{k=0}^{\infty} \gamma^k R_{t+k}.
\]
The specific structure of \(R_t\) is designed in the next section to align with the trade-off between detection, communication quality, and sensing effort.
\section{Proposed Method}
\label{sec:Proposed_Method_Content}

Our proposed solution to optimize beamforming and sensing is an advanced DRL agent built upon the PPO algorithm. PPO is a state-of-the-art, policy-based method known for its stability and robust performance in complex, high-dimensional environments. The agent is trained to dynamically adjust the base station's beam azimuth and ISAC effort to maximize security while maintaining communication quality. To overcome the significant exploration challenges inherent in this task, we introduce an intensive curriculum learning phase. 

\subsection{PPO Agent Design}
The agent employs an Actor-Critic architecture, which consists of two separate neural networks:
\begin{itemize}
    \item \textbf{The Actor Network:} This network learns the policy \(\pi(a_t|s_t)\) by taking the current state as the input and outputting a probability distribution over the discrete action space.
    \item \textbf{The Critic Network:} This network estimates the value function \(V(s_t)\), which predicts the expected cumulative reward from a given state. This value is used to assess the quality of the Actor's actions.
\end{itemize}

Both networks are implemented as fully connected multilayer perceptrons (MLPs). The input layer corresponds to the 7-dimensional state vector. This is followed by two hidden layers with 256 and 128 neurons, respectively, using ReLU activation functions. The Actor network has a final output layer with 5 neurons and a softmax activation to represent the action probabilities, while the Critic network has a single linear output neurons for the state value. The models are optimized using the Adam optimizer. PPO is chosen for its stability and efficiency in continuous and noisy environments. Unlike DQN, which requires discrete actions, PPO handles continuous control directly. Compared to A2C, PPO’s clipping improves training stability, which is crucial in sparse-reward tasks like ours. Imitation Learning is unsuitable here due to the lack of expert demonstrations.
\subsection{Reward Design}
\label{subsec:Reward_Design}
The reward function, \(R_t\), at each step \(t\), is designed as a weighted sum of key behavioral indicators to balance proactive defense and communication quality:
\begin{equation}
\label{eq:reward_function}
\begin{split}
    R_t = & \quad w_{\text{det}} \cdot \mathbb{I}(\text{conf}_t > 0.7) \\
          & + w_{\text{pro}} \cdot \mathbb{I}(\text{range}_t^{\text{true}} < 80\text{m} \land \text{effort}_t > 0.8) \\
          & - w_{\text{unaware}} \cdot \mathbb{I}(\text{range}_t^{\text{true}} < 80\text{m} \land \text{conf}_t < 0.7) \\
          & + w_{\text{com}} \cdot \mathbb{I}(\text{SINR}_t > 5\text{dB}),
\end{split}
\end{equation}
where \(\mathbb{I}(\cdot)\) is an indicator function. The weights prioritize security: \(w_{\text{det}} = 150\), \(w_{\text{pro}} = 25\), \(w_{\text{unaware}} = 5\), and \(w_{\text{com}} = 0.5\). This structure encourages confident detections, rewards proactive sensing, penalizes ignorance of close threats, and preserves minimum SINR for service quality.

\subsection{Training with Intensive Curriculum Learning}
A primary challenge in this problem is the vast and sparse reward landscape where the agent may 
not find the high-reward states corresponding to a successful detection. Initial experiments without a curriculum confirm this, resulting in a 0\% detection rate as the agent settles in a suboptimal policy of only maximizing the SINR. We design an intensive, two-phase curriculum learning strategy to solve this issue.

\subsubsection{Phase 1: Forced Success Curriculum (First 1500 Episodes)}
The goal of this initial phase is to guarantee that the agent experiences successful detection, thereby learning the value of security-oriented actions. In each of the first 1500 training episodes, a "forced success" mechanism selects five unique, random time steps and overrides the agent's actions to set the beam azimuth to the attacker's true direction and ISAC effort to its maximum value of 1.0, ensuring 100\% guaranteed detection and associating the large 
reward of 150 with a specific state-action context. This initial phase, spanning 1500 episodes, exposes the agent to a total of 7,500 guaranteed successful detection experiences, 5 per episode. This dense exposure proves to be critical in seeding the agent's memory with high-value experiences, enabling it to overcome the exploration challenge.

\subsubsection{Phase 2: Autonomous Learning with Guided Exploration (Post-Curriculum)}
The curriculum ends after 1500 episodes, and the agent becomes fully autonomous. Now equipped with the knowledge that a high-reward security strategy exists, it has the necessary foundation and motivation to explore and refine this strategy on its own. We employ a guided exploration mechanism during this phase to prevent catastrophic forgetting and reinforce the learned behavior. At each step, there is a small (10\%) probability that the environment will override the agent's action and instead execute the "forced success" action. This intermittent guidance ensures the agent remains focused on the effective security policy while still allowing it to learn the complex trade-offs defined by the reward function.

\section{Results}
\label{sec:Results}
\begin{table*}[t]
\centering
\caption{Detailed statistical performance comparison of the final PPO agent and the physics-based SecBeam baseline.}
\label{tab:performance_comparison_final}
\begin{tabular}{lcc}
\toprule
\textbf{Metric} & \textbf{Baseline (SecBeam Protocol)} & \textbf{Final PPO Agent (Ours)} \\
\midrule
\multicolumn{3}{l}{\textit{\textbf{Communication Performance}}} \\
Mean SINR (dB) & 26.80 & 13.10 \\
Std. Dev. of SINR (dB) & 21.49 & 11.32 \\
Median SINR (dB) & 32.25 & 16.47 \\
Min / Max SINR (dB) & -26.00 / 56.80 & -26.00 / 27.49 \\
\midrule
\multicolumn{3}{l}{\textit{\textbf{Security Performance}}} \\
Mean Detection Rate (\%) & \textbf{68.00\%} & \textbf{92.80\%} \\
Std. Dev. of Detection Rate (\%) & 46.65\% & 13.51\% \\
Median Detection Rate (\%) & 100.00\% & \textbf{100.00\%} \\
Max Detection Rate in an Episode (\%) & 100.00\% & 100.00\% \\
\midrule
\multicolumn{3}{l}{\textit{\textbf{Reward Statistics}}} \\
Mean Cumulative Reward & N/A & 1976.60 \\
Std. Dev. of Reward (Stability) & N/A & 1740.42 \\
Median Cumulative Reward & N/A & 1420.00 \\
Min / Max Reward & N/A & 20.00 / 8750.00 \\
\bottomrule
\end{tabular}
\end{table*}

\subsection{Simulation and Implementation Setup}

The framework is implemented in Python using TensorFlow~2.17, with the mmWave channel simulated via the Sionna library \cite{sionna} configured for the 3GPP TR 38.901 Urban Macrocell (UMa) model. The full implementation is available online\footnote{\url{https://github.com/CLIS-WPI/Secure-Beamforming}}.

The base station is equipped with a uniform planar array (UPA) of $8 \times 8 = 64$ vertically polarized antenna elements. It communicates with a legitimate user equipment (UE) and simultaneously attempts to detect a beam-stealing attacker. Both UE and attacker are equipped with single-antenna omnidirectional receivers. The carrier frequency is set to 28~GHz and the total available bandwidth is 100~MHz. The total transmit power is 30~dBm, and the noise power spectral density is fixed at $-174$~dBm/Hz.

To simulate realistic dynamics, small positional perturbations are added as Gaussian noise in each episode to simulate mobility. The sensing subsystem introduces detection errors modeled as Gaussian noise, with a standard deviation of 3 degrees in azimuth and 1.5 meters in range.

The PPO agent observes a 7-dimensional state vector consisting of the user's SINR, the current beam azimuth, the estimated azimuth and range of the attacker, detection confidence, and the ground truth location of the attacker (used only during training for reward computation).

The key hyperparameters of the PPO agent are summarized in Table~\ref{tab:hyperparameters}.
\begin{table}[h]
\centering
\caption{PPO Agent Hyperparameters}
\label{tab:hyperparameters}
\begin{tabular}{lc}
\toprule
\textbf{Hyperparameter} & \textbf{Value} \\
\midrule
Actor Learning Rate & $3 \times 10^{-4}$ \\
Critic Learning Rate & $1 \times 10^{-3}$ \\
Discount Factor ($\gamma$) & 0.99 \\
GAE Lambda ($\lambda$) & 0.95 \\
PPO Clip Epsilon & 0.2 \\
Training Epochs ($K$) & 40 \\
Batch Size & 4096 \\
\bottomrule
\end{tabular}
\end{table}

\subsection{Overall Performance Evaluation}
The primary outcome of our framework is the agent's ability to learn a highly effective, security-first policy while maintaining excellent communication quality. As summarized in Table~\ref{tab:performance_comparison_final}, our PPO agent achieves a mean attacker detection rate of 92.80\%. The median detection rate is 100\%, indicating that in over half of all autonomous episodes, the agent successfully detects the attacker in every single step. This high median value underscores the reliability of the learned policy, demonstrating its consistent success once the security-oriented strategy is triggered.

We compare its performance against a physics-based defense protocol, SecBeam \cite{10735492}, with full results in Table~\ref{tab:performance_comparison_final}. The SecBeam baseline achieves a higher mean SINR (26.80 dB) but proves less reliable, with a mean detection rate of only 68.00\%. In contrast, our PPO agent achieves a vastly superior 92.80\% detection rate by learning a more effective trade-off, maintaining a robust average SINR of 13.10 dB. This adaptive balancing of security and communication quality is a fundamental advantage of our DRL-based approach over non-adaptive mechanisms.

\begin{figure*}[t]
    \centering
    \includegraphics[width=2.0\columnwidth]{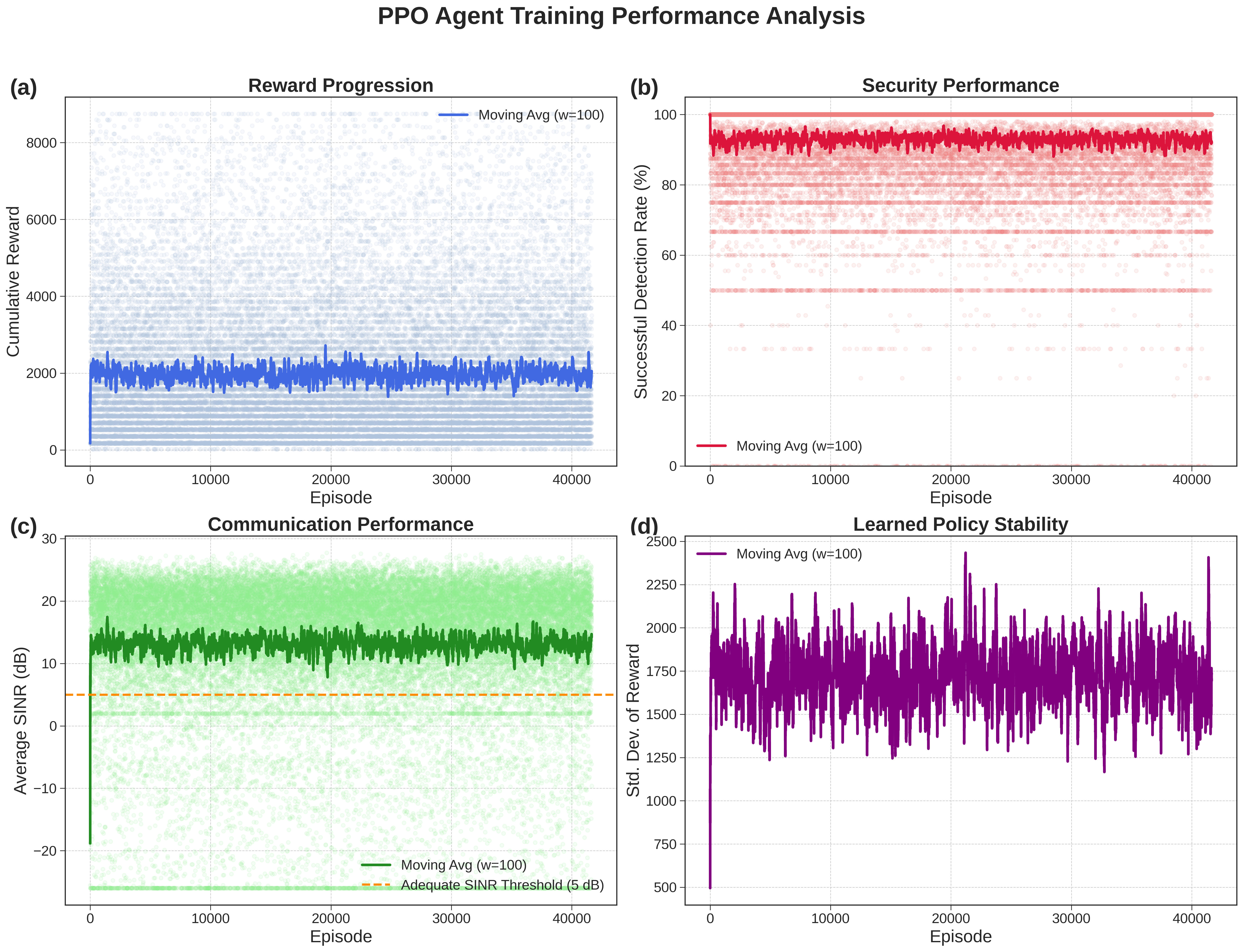}
    \caption{Training performance over simulation time, showing the agent's cumulative reward (a), the successful detection rate (b), the user's average SINR (c), and the standard deviation of the reward as a measure of policy stability (d). The agent shows strong convergence after the initial 1500-episode curriculum phase.}
    \label{fig:training_curves}
\end{figure*}

Figure~\ref{fig:training_curves} illustrates the agent's successful convergence to an effective policy following the curriculum phase. Fig.~\ref{fig:training_curves}(a) shows the positive reward trend. Fig.~\ref{fig:training_curves}(b) illustrates that the detection rate nears 100\% after the curriculum phase, indicating the agent's successful escape from the initial exploration trap and convergence to a highly rewarding policy. Fig.~\ref{fig:training_curves}(c) shows that this security is achieved while consistently maintaining a high-quality SINR above the 5~dB threshold. Fig.~\ref{fig:training_curves}(d) illustrates the stability of the learned policy, where the standard deviation of the reward, although high because of the agent's adaptive nature, remains stable after the initial learning phase.

\subsection{Analysis of the Learned Adaptive Policy}
\begin{figure}[h!]
    \centering
    \includegraphics[width=1.0\columnwidth]{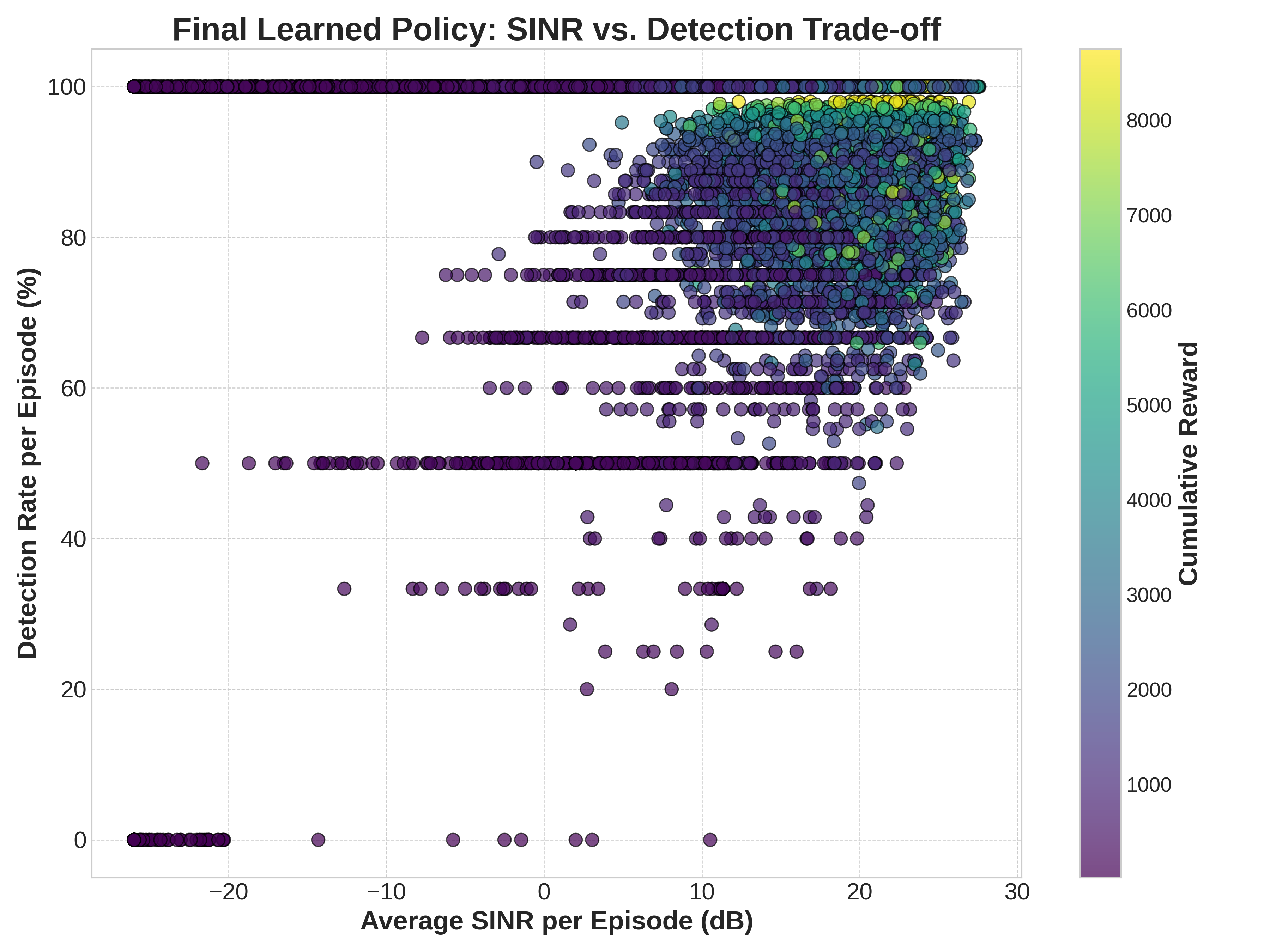}
    \caption{A scatter plot visualizing the final learned policy. Each point represents an episode, showing the trade-off between the achieved SINR and detection rate, with color indicating the total reward.}
    \label{fig:tradeoff_plot}
\end{figure}

A key finding is that the agent learns an adaptive, intelligent policy rather than a simple, static one. This is evidenced by the high standard deviation of the final reward (1740.42), which reflects the agent's ability to tailor its strategy to the specific, dynamic conditions of each episode. Fig.~\ref{fig:tradeoff_plot} visualizes this trade-off. The plot shows that the agent can achieve high detection rates across a wide range of SINR values. Episodes with lower SINR often correspond to scenarios where the agent must take more aggressive beamforming actions to secure the link, sacrificing some communication quality for near-perfect security. Conversely, in scenarios where the threat is less immediate, it can achieve both high detection rates and excellent SINR. This dynamic balancing act is the hallmark of an intelligent defense system. The high variance in rewards is therefore not a sign of instability, but rather a direct consequence of this intelligent, state-dependent adaptability. This adaptive behavior is further confirmed by the agent's resource allocation strategy, shown in Fig.~\ref{fig:isac_strategy}, where the ISAC effort is decisively increased only when a threat is perceived as near.

\subsection{Analysis of the ISAC Effort Strategy}
\begin{figure}[h!]
    \centering
    \includegraphics[width=1.0\columnwidth]{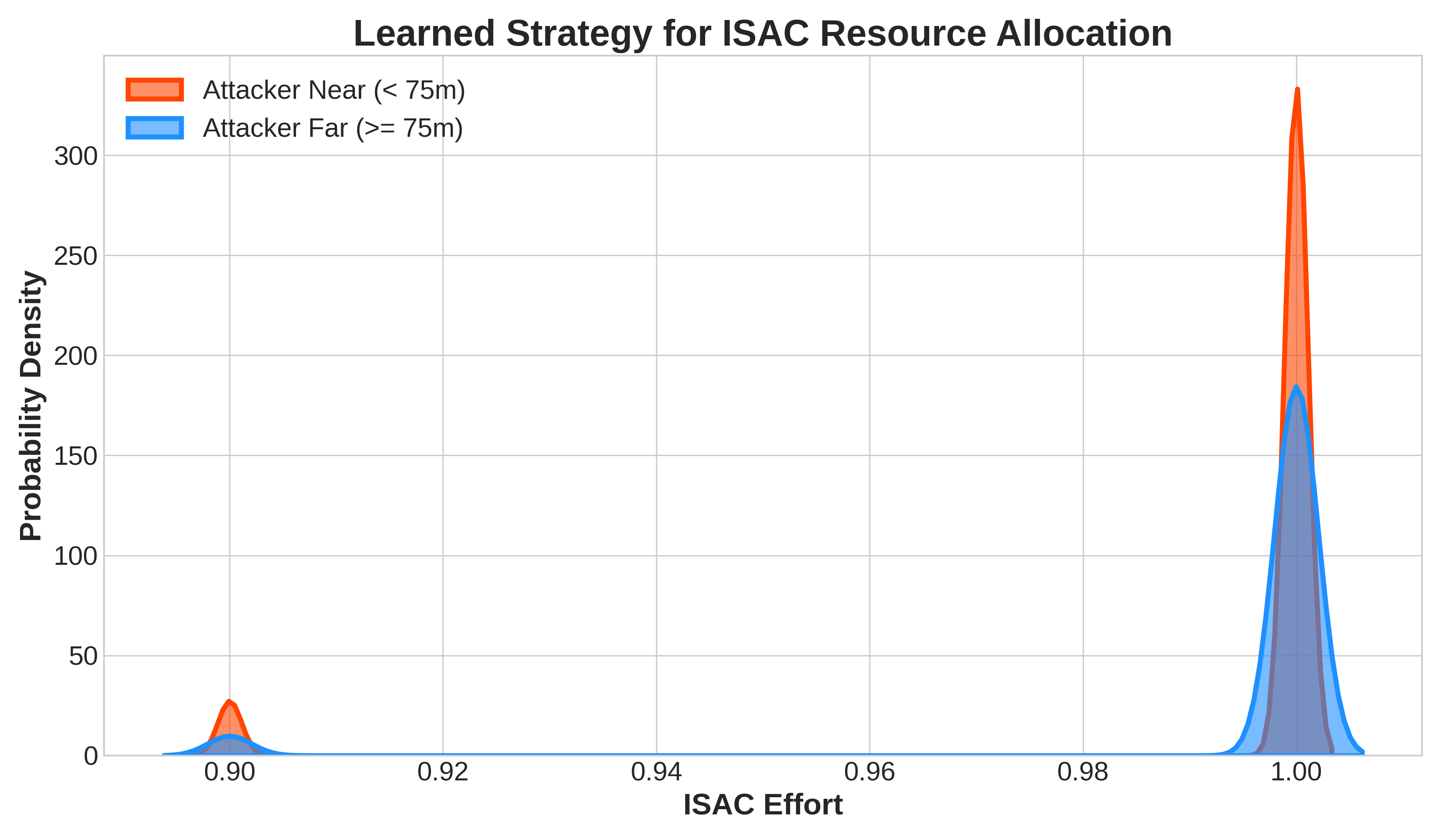} 
    \caption{The learned ISAC resource allocation strategy. The agent distinguishes between near and far threats, allocating maximum effort only when the attacker is close.}
    \label{fig:isac_strategy}
\end{figure}

We analyze the ISAC effort decisions based on threat proximity to validate that the agent learns a resource-efficient policy. Fig.~\ref{fig:isac_strategy} shows that the agent learns a distinct, bimodal strategy. When the attacker is near (< 75m), the agent overwhelmingly allocates the maximum ISAC effort (a sharp peak at 1.0) to ensure detection. However, when the attacker is far away, it predominantly uses a lower effort, conserving resources. This targeted intensification of sensing demonstrates that the agent has learned to manage its ISAC resources efficiently based on the immediate security context, rather than employing a naive, always-on sensing strategy.

\section{Discussion}
\label{sec:Discussion}

Our DRL agent, trained with intensive curriculum learning, develops an effective and intelligent policy that balances threat detection with communication efficiency. It achieves a 92.8\% mean attacker detection rate and maintains a 13.1 dB average SINR, with a 100\% median detection rate, indicating reliable threat identification in over half of the autonomous episodes.

The learned policy demonstrates a nuanced and resource-efficient behavior. Its highly adaptive nature is reflected in the high standard deviation of the reward values. As discussed in Section~\ref{sec:Results}, this variability is not indicative of instability but rather results from an intelligent decision-making process that dynamically adapts to the specific conditions of each scenario. This contrasts sharply with the rigid, rule-based logic of the SecBeam baseline, whose performance variance arises from binary outcomes rather than a learned, context-aware strategy.

The analysis of the ISAC effort strategy (Fig.~\ref{fig:isac_strategy}) confirms this intelligent behavior. The agent learns to intensify sensing efforts primarily when the perceived threat is near and justifies the resource cost, a sophisticated, security-driven use of ISAC that distinguishes our work from general purpose sensing applications\cite{10812728,Zhang2024}.

While the framework is highly effective, its limitations warrant discussion. The current model is limited to a single-attacker context, and its performance inherently depends on the fidelity of ISAC sensing data. Although our curriculum learning proves effective, significant sensor inaccuracies could still mislead the agent's decisions. Finally, translating simulated results to hardware presents challenges, as delays and imperfect CSI could degrade sensing and communication performance. These challenges mark avenues for future investigation.

\section{Conclusion and Future Work}
\label{sec:Conclusion}

This paper presents a PPO-based DRL framework to secure mmWave communications against beam-stealing attacks, using intensive curriculum learning to overcome exploration challenges and achieve a robust policy. Our agent attains a mean detection rate of 92.8\%, significantly outperforming the 68\% of the physics-based SecBeam baseline, by learning to intelligently balance security and communication efficiency. The approach assumes a single, non-adaptive attacker and perfect CSI, with future work focusing on multi-attacker scenarios, imperfect CSI, and pursuing real-world testbed validation.

\section*{\textcolor{black}{Acknowledgment}}
{\small
\noindent
This material is based upon work supported in part by NSF under Awards CNS-2120442 and IIS-2325863, and NTIA under Award No. 51-60-IF007. Any opinions, findings, and conclusions or recommendations expressed in this publication are those of the author(s) and do not necessarily reflect the views of the NSF and NTIA.
}

{\small
\bibliographystyle{IEEEtran}
\bibliography{bib/main}

\begin{thebibliography}{10}
\providecommand{\url}[1]{#1}
\csname url@samestyle\endcsname
\providecommand{\newblock}{\relax}
\providecommand{\bibinfo}[2]{#2}
\providecommand{\BIBentrySTDinterwordspacing}{\spaceskip=0pt\relax}
\providecommand{\BIBentryALTinterwordstretchfactor}{4}
\providecommand{\BIBentryALTinterwordspacing}{\spaceskip=\fontdimen2\font plus
\BIBentryALTinterwordstretchfactor\fontdimen3\font minus \fontdimen4\font\relax}
\providecommand{\BIBforeignlanguage}[2]{{%
\expandafter\ifx\csname l@#1\endcsname\relax
\typeout{** WARNING: IEEEtran.bst: No hyphenation pattern has been}%
\typeout{** loaded for the language `#1'. Using the pattern for}%
\typeout{** the default language instead.}%
\else
\language=\csname l@#1\endcsname
\fi
#2}}
\providecommand{\BIBdecl}{\relax}
\BIBdecl

\bibitem{10422712}
Q.~Xue, C.~Ji, S.~Ma, J.~Guo, Y.~Xu, Q.~Chen, and W.~Zhang, ``A survey of beam management for mmwave and {THz} communications towards {6G},'' \emph{IEEE Communications Surveys \& Tutorials}, vol.~26, no.~3, pp. 1520--1559, 2024.

\bibitem{10735492}
J.~Li, L.~Lazos, and M.~Li, ``Secbeam: Securing mmwave beam alignment against beam-stealing attacks,'' in \emph{2024 IEEE Conference on Communications and Network Security (CNS)}, 2024, pp. 1--9.

\bibitem{10.1145/3212480.3212499}
D.~Steinmetzer, Y.~Yuan, and M.~Hollick, ``Beam-stealing: Intercepting the sector sweep to launch man-in-the-middle attacks on wireless ieee 802.11ad networks,'' in \emph{Proceedings of the 11th ACM Conference on Security \& Privacy in Wireless and Mobile Networks}, ser. WiSec '18.\hskip 1em plus 0.5em minus 0.4em\relax New York, NY, USA: Association for Computing Machinery, 2018, p. 12–22.

\bibitem{10536135}
N.~González-Prelcic, M.~Furkan~Keskin, O.~Kaltiokallio, M.~Valkama, D.~Dardari, X.~Shen, Y.~Shen, M.~Bayraktar, and H.~Wymeersch, ``The integrated sensing and communication revolution for {6G}: Vision, techniques, and applications,'' \emph{Proceedings of the IEEE}, vol. 112, no.~7, pp. 676--723, 2024.

\bibitem{Yang2022}
Y.~Yang, X.~Wei, R.~Xu, W.~Wang, L.~Peng, and Y.~Wang, ``Jointly beam stealing attackers detection and localization without training: an image processing viewpoint,'' \emph{Frontiers of Computer Science}, vol.~17, no.~3, p. 173704, 2022.

\bibitem{10.1145/3636534.3690669}
M.~Xu, Y.~He, X.~Li, J.~Hu, Z.~Chen, F.~Xiao, and J.~Luo, ``Beamforming made malicious: Manipulating wi-fi traffic via beamforming feedback forgery,'' in \emph{Proceedings of the 30th Annual International Conference on Mobile Computing and Networking}, ser. ACM MobiCom '24.\hskip 1em plus 0.5em minus 0.4em\relax New York, NY, USA: Association for Computing Machinery, 2024, p. 908–922.

\bibitem{10026317}
B.~Qiu, W.~Cheng, and W.~Zhang, ``Robust multi-beam secure mmwave wireless communication for hybrid wiretapping systems,'' \emph{IEEE Transactions on Information Forensics and Security}, vol.~18, pp. 1393--1406, 2023.

\bibitem{xu2025synergizingcoverttransmissionmmwave}
L.~Xu, B.~Wang, and Z.~Cheng, ``Synergizing covert transmission and mmwave isac for secure iot systems,'' 2025.

\bibitem{Kuzlu2023}
M.~Kuzlu, F.~O. Catak, U.~Cali, E.~Catak, and O.~Guler, ``Adversarial security mitigations of mmwave beamforming prediction models using defensive distillation and adversarial retraining,'' \emph{International Journal of Information Security}, vol.~22, no.~2, pp. 319--332, Apr. 2023.

\bibitem{natanzi2025onlinelearningbasedadaptivebeam}
S.~B.~H. Natanzi, Z.~Zhu, and B.~Tang, ``Online learning-based adaptive beam switching for 6g networks: Enhancing efficiency and resilience,'' 2025.

\bibitem{10773704}
H.~Mohammadi, M.~Zhang, A.~Jha, V.~Marojevic, R.~Chou, and T.~Kim, ``Fortifying 5g networks: Defending against jamming attacks with multipath communications,'' in \emph{MILCOM 2024 - 2024 IEEE Military Communications Conference (MILCOM)}, 2024, pp. 680--681.

\bibitem{sionna}
J.~Hoydis, S.~Cammerer, F.~Ait~Aoudia, M.~Nimier-David, L.~Maggi, G.~Marcus, A.~Vem, and A.~Keller, ``Sionna: A gpu-accelerated library for link-level simulations,'' 2022, version 1.1.0, NVIDIA, available from NVIDIA Developer Zone.

\bibitem{10812728}
D.~Wen, Y.~Zhou, X.~Li, Y.~Shi, K.~Huang, and K.~B. Letaief, ``A survey on integrated sensing, communication, and computation,'' \emph{IEEE Communications Surveys \& Tutorials}, pp. 1--1, 2024.

\bibitem{Zhang2024}
J.~Zhang, W.~Lu, C.~Xing, N.~Zhao, N.~Al-Dhahir, G.~K. Karagiannidis, and X.~Yang, ``Intelligent integrated sensing and communication: a survey,'' \emph{Science China Information Sciences}, vol.~68, no.~3, p. 131301, December 2024.

\end{thebibliography}
}

\end{document}